\documentclass[%
 reprint,
superscriptaddress,
showpacs,
 amsmath,amssymb,
 aps,
pra,
]{revtex4-1}

\usepackage{graphicx}% Include figure files
\usepackage{dcolumn}% Align table columns on decimal point
\usepackage{bm}% bold math
\newcommand{\bra}[1]{\ensuremath{\left\langle#1\right|}}
\newcommand{\ket}[1]{\ensuremath{\left|#1\right\rangle}}

\begin{document}

\preprint{APS/123-QED}

\title{Efficient preparation and detection of microwave dressed-state qubits and qutrits with trapped ions}% Force line breaks with \\

\author{J. Randall}
\affiliation{Department of Physics and Astronomy, University of Sussex, Brighton, BN1 9QH, UK}
\affiliation{QOLS, Blackett Laboratory, Imperial College London, London, SW7 2BW, UK}
\author{S. Weidt}
\affiliation{Department of Physics and Astronomy, University of Sussex, Brighton, BN1 9QH, UK}
\author{E. D. Standing}
\affiliation{Department of Physics and Astronomy, University of Sussex, Brighton, BN1 9QH, UK}
\author{K. Lake}
\affiliation{Department of Physics and Astronomy, University of Sussex, Brighton, BN1 9QH, UK}
\author{S. C. Webster}
\affiliation{Department of Physics and Astronomy, University of Sussex, Brighton, BN1 9QH, UK}
\author{D. F. Murgia}
\affiliation{Department of Physics and Astronomy, University of Sussex, Brighton, BN1 9QH, UK}
\affiliation{QOLS, Blackett Laboratory, Imperial College London, London, SW7 2BW, UK}
\author{T. Navickas}
\affiliation{Department of Physics and Astronomy, University of Sussex, Brighton, BN1 9QH, UK}
\author{K. Roth}
\affiliation{Department of Physics and Astronomy, University of Sussex, Brighton, BN1 9QH, UK}
\author{W. K. Hensinger}
\email{w.k.hensinger@sussex.ac.uk}
\affiliation{Department of Physics and Astronomy, University of Sussex, Brighton, BN1 9QH, UK}

\date{\today}% It is always \today, today,

\begin{abstract}
We demonstrate a method for preparing and detecting all eigenstates of a three-level microwave dressed system with a single trapped ion. The method significantly reduces the experimental complexity of gate operations with dressed-state qubits, as well as allowing all three of the dressed-states to be prepared and detected, thereby providing access to a qutrit that is well protected from magnetic field noise. In addition, we demonstrate individual addressing of the clock transitions in two ions using a strong static magnetic field gradient, showing that our method can be used to prepare and detect microwave dressed-states in a string of ions when performing multi-ion quantum operations with microwave and radio frequency fields. The individual addressability of clock transitions could also allow for the control of pairwise interaction strengths between arbitrary ions in a string using lasers.
\end{abstract}

\maketitle

%\tableofcontents

\section{Introduction}

Preparation and detection of quantum states is essential for any quantum information processor. Qubits encoded in the hyperfine ground states of trapped ions can be prepared using optical pumping and detected using state-dependent fluorescence, both with high fidelity \cite{Noek, Burrell, Harty}. Instead of encoding the qubit in the bare atomic states, it can be advantageous to instead encode quantum information in the states formed when the ion is dressed by continuously applied microwave or laser fields, known as dressed-states \cite{Timoney, Webster, Cohen, Bermudez2, Tan, Navon, Rabl}. These states are less sensitive to dephasing from magnetic field noise and hence their use can increase the coherence time of the system by several orders of magnitude. However, preparing and detecting dressed-states requires additional manipulation which can lead to decoherence, and therefore it is important for robust and scalable methods to be developed. 

One instance in which the use of dressed-states has so far been shown to be beneficial is for performing high-fidelity gates using microwave and radio frequency (RF) fields \cite{Timoney, Webster}. One method of performing such gates requires a static magnetic field gradient to be applied to a chain of ions \cite{Mintert1}. This produces a coupling between the spin and motional states of the ions, which can be used for multi-qubit gates, as well as allowing for individual addressing of ions in frequency space \cite{Johanning2,Khromova1,Piltz2}. One drawback, however, is that the scheme requires states with different magnetic moments to be used as qubit states, and therefore the qubits are highly sensitive to magnetic field noise. The particular scheme demonstrated in \cite{Timoney, Webster} has shown how microwave dressed-states can protect against such noise. Here, a single $^{171}\text{Yb}^+$ ion was used, for which two microwave fields dress three atomic states, and RF fields can be used to manipulate a qubit formed of one of these dressed-states and a fourth magnetic field insensitive state. In order to prepare the required dressed-state the amplitudes of the microwave dressing fields are slowly modulated in a Stimulated Raman Adiabatic Passage (STIRAP) process, along with additional microwave pulses \cite{Timoney, Webster}. While such a process does allow for the preparation of one of the dressed-states, it does not easily lend itself to preparing all three dressed-states directly and involves the use of magnetic field sensitive transitions, which can limit the achievable fidelity. 

In this work, we demonstrate a novel method for preparing and detecting all three of the dressed-states obtained by dressing three bare-states in a single $^{171}\text{Yb}^+$ ion with microwave fields.
The three dressed-states could be used to form a qutrit which is well protected from magnetic field noise. Qutrits can have advantages over qubits in various applications, including faster quantum information processing \cite{Klimov, Ralph, Bartlett, Lanyon2}, the study of entanglement in higher dimensional systems \cite{Collins, Kaszlikowski}, quantum simulations of spin-1 systems \cite{Cohen, Senko} and more robust quantum cryptography protocols \cite{Bruss, Cerf}. Preparation and detection of a dressed-state qutrit is therefore an important step towards the realisation of such experiments in trapped ions.
Furthermore, by utilising a clock transition we eliminate the use of magnetic field sensitive transitions during the preparation and detection sequence, as well as the requirement to modulate the amplitude of the dressing fields. This eases experimental requirements for achieving high preparation and detection fidelities of the dressed system.

The manuscript is arranged in the following way. We begin in section \ref{sec:dressedstates} by reviewing the microwave dressed-state scheme from Refs. \cite{Timoney, Webster}, which is also used in the following work, and show how it protects against dephasing from magnetic field fluctuations. We then briefly outline the STIRAP preparation and detection method in section \ref{sec:prepdet}, before introducing our new method and demonstrating that all three dressed-states can be prepared and detected. Finally, in section \ref{sec:clock}, we show how our method can be integrated into the static magnetic field gradient quantum logic scheme \cite{Mintert1} by demonstrating individual addressing of the clock transitions in two ions using a magnetic field gradient.

\section{Microwave dressed-states}\label{sec:dressedstates}
\begin{figure}
\centering
\includegraphics[width=\columnwidth]{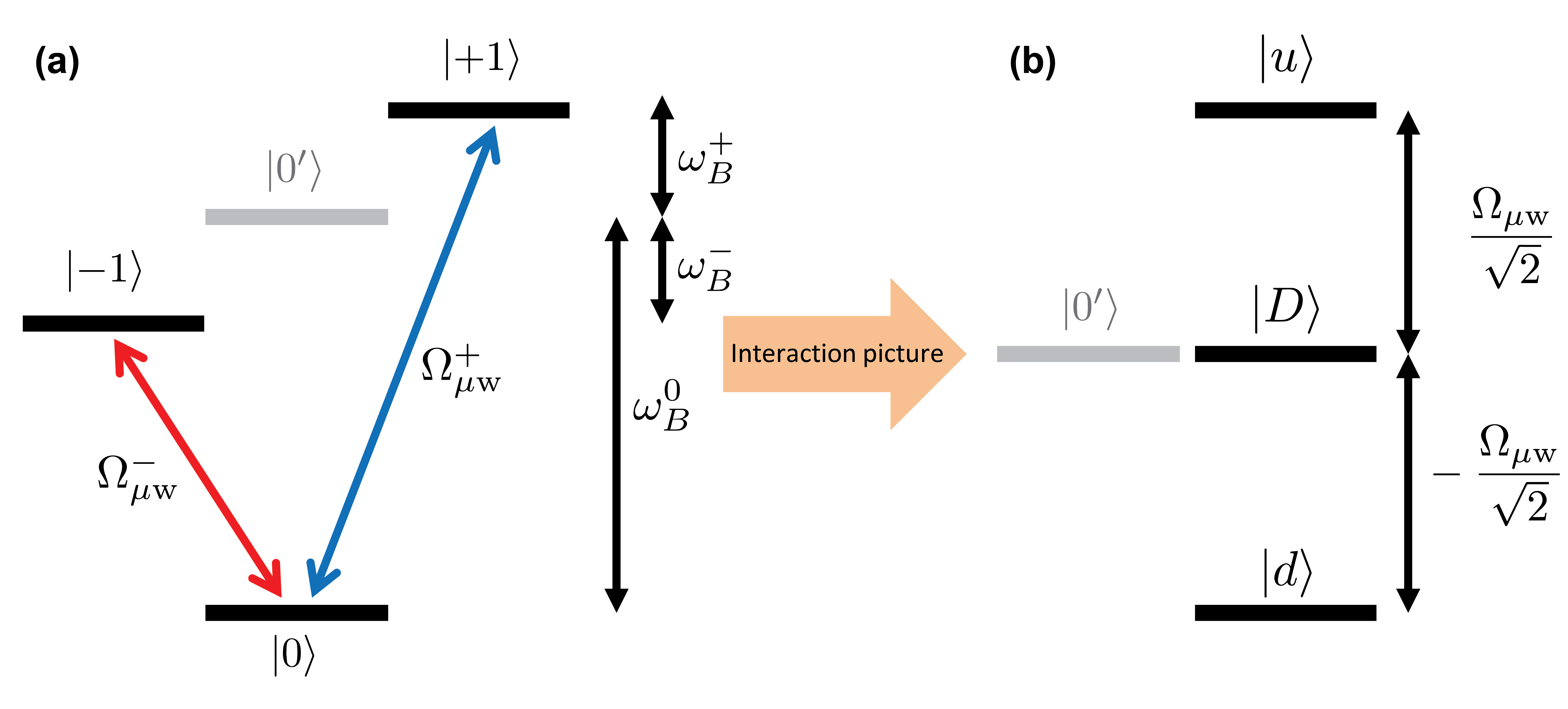}
\caption{(a) Energy level diagram of the $^{2}$S$_{1/2}$ ground state hyperfine manifold of $^{171}$Yb$^{+}$ where the degeneracy of the Zeeman states has been lifted by a static magnetic field. Here the transitions $\ket{0}\leftrightarrow\ket{+1}$ and $\ket{0}\leftrightarrow\ket{-1}$ are coupled with resonant microwave radiation to obtain the three dressed-states $\ket{d}$, $\ket{D}$ and $\ket{u}$. (b) The resultant energy level diagram in the dressed basis when the Rabi frequencies of the dressing fields are set to be equal ($\Omega_{\mu\text{w}}^+ = \Omega_{\mu\text{w}}^- = \Omega_{\mu\text{w}}$). The energy difference between the dressed-states in this case is given by $\pm\Omega_{\mu\text{w}}/\sqrt{2}$.}
\label{dressed_level_diagram}
\end{figure}

A dressed-state is an eigenstate of the Hamiltonian which describes an atomic system being driven by resonant electromagnetic fields. In particular, microwave dressed-states can be created in the $^{2}$S$_{1/2}$ ground state hyperfine manifold of $^{171}$Yb$^{+}$ \cite{Timoney, Webster}. This manifold consists of four states, of which $\ket{0} \equiv$ $^{2}$S$_{1/2}\ket{F=0}$ and $\ket{0'} \equiv$ $^{2}$S$_{1/2}\ket{F=1, m_F = 0}$ are insensitive to magnetic fields at low field stengths, while $\ket{+1} \equiv$ $^{2}$S$_{1/2}\ket{F=1, m_{F}=+1}$ and $\ket{-1} \equiv$ $^{2}$S$_{1/2}\ket{F=0, m_{F}=-1}$ are magnetic field sensitive. To make use of the magnetic field gradient scheme in Ref. \cite{Mintert1}, at least one of the magnetic field sensitive states must be used. The transition frequencies, which correspond to those shown in Fig. \ref{dressed_level_diagram} (a), can be derived from the Breit-Rabi formula to be \cite{Breit}
\begin{equation}\label{breitrabi}\begin{split}
\omega_B^{+} &= \frac{\omega_{0}}{2}\big(1 + \chi - \sqrt{1+\chi^2}\big) \\
\omega_B^{-} &= -\frac{\omega_{0}}{2}\big(1 - \chi - \sqrt{1+\chi^2}\big) \\
\omega_B^0 &= \omega_{0}\sqrt{1+\chi^2},
\end{split}
\end{equation}
where $\chi = g_J \mu_B B / \hbar \omega_0$ and we have neglected the contribution from the nuclear spin as the nuclear magneton $\mu_N$ is much smaller than the Bohr magneton $\mu_B$. Here $\omega_0/2\pi\simeq12{,}642{,}812.1$ kHz \cite{Fisk} is the zero field transition frequency between the F = 0 and F = 1 levels, $g_J$ is the electronic g-factor and $B$ is the applied magnetic field. In all the following work, we work in the low magnetic field regime where $\chi\ll 1$.

At low magnetic field strengths $\partial_B\omega_B^+$ and $\partial_B\omega_B^-$ are approximately constant, so the sensitivity of qubits based on these transitions to magnetic field noise is independent of the mean magnetic field. At non-zero fields, the $\ket{0}\leftrightarrow\ket{0'}$ clock transition becomes first-order sensitive to magnetic fields. However, it retains a much smaller sensitivity to magnetic field fluctuations, of size $2\chi$ relative to the magnetic field sensitive transitions involving the states $\ket{+1}$ and $\ket{-1}$.

To create the three-level dressed system, the magnetic field sensitive states $\ket{+1}$ and $\ket{-1}$ are coupled to the magnetic field insensitive state $\ket{0}$ using two resonant microwave fields with Rabi frequencies $\Omega_{\mu\text{w}}^+$ and $\Omega_{\mu\text{w}}^-$ respectively, as shown in Fig. \ref{dressed_level_diagram} (a).

Setting $\Omega_{\mu\text{w}}^+ = \Omega_{\mu\text{w}}^- = \Omega_{\mu\text{w}}$, moving to the interaction picture, and performing the rotating wave approximation (RWA), the Hamiltonian describing this system is given by
\begin{equation}\label{mw_int_hamiltonian}
\hat{H}_{\mu\text{w}}=\frac{\hbar\Omega_{\mu\text{w}}}{2}\bigg(\ket{+1}\bra{0}+\ket{-1}\bra{0}+\text{H.c.}\bigg),
\end{equation}
where H.c stands for Hermitian conjugate and we have defined the phases of the microwave fields to be zero for convenience. This Hamiltonian has three eigenstates, or dressed-states, given by
\begin{equation}\label{dressedstates}\begin{split}
\ket{D}&=\frac{1}{\sqrt{2}}\big(\ket{+1}-\ket{-1}\big) \\
\ket{u}&=\frac{1}{2}\ket{+1}+\frac{1}{2}\ket{-1}+\frac{1}{\sqrt{2}}\ket{0} \\
\ket{d}&=\frac{1}{2}\ket{+1}+\frac{1}{2}\ket{-1}-\frac{1}{\sqrt{2}}\ket{0},
\end{split}\end{equation}
\noindent
and can then be written in terms of these dressed-states as
\begin{equation}\label{dressed_mw_hamiltonian}
\hat{H}_{\mu\text{w}}=\frac{\hbar\Omega_{\mu\text{w}}}{\sqrt{2}}\bigg(\ket{u}\bra{u}-\ket{d}\bra{d}\bigg).
\end{equation}
\noindent
The energies of the dressed-states $\ket{u}$ and $\ket{d}$ are separated from the dressed-state $\ket{D}$ by $\hbar\Omega_{\mu\text{w}}/\sqrt{2}$ and $-\hbar\Omega_{\mu\text{w}}/\sqrt{2}$ respectively. The resultant energy level diagram is shown in Fig. \ref{dressed_level_diagram} (b). 

An important feature of these dressed-states is that they are robust to magnetic field fluctuations. This can be seen by treating the magnetic field fluctuations as a perturbation of the form
\begin{equation}\label{magnetic_field_perturbation}
\hat{H}_{\text{p}}=\hbar\lambda_{0}(t)\bigg(\ket{+1}\bra{+1}-\ket{-1}\bra{-1}\bigg),
\end{equation}
where $\lambda_0(t)$ is an arbitrary time dependent function. In the dressed basis, Eq. \ref{magnetic_field_perturbation} becomes
\begin{equation}\label{dressed_magnetic_field_perturbation}
\hat{H}_{\text{p}}=\frac{\hbar\lambda_{0}(t)}{\sqrt{2}}\bigg(\ket{D}\bra{u}+\ket{D}\bra{d}+\text{H.c.}\bigg).
\end{equation}
\noindent
Magnetic field fluctuations will therefore try to drive population between $\ket{D}$, $\ket{u}$ and $\ket{d}$, but these states are separated by an energy gap of $\Omega_{\mu\text{w}}/\sqrt{2}$. Consequently, only magnetic field fluctuations with a frequency at or near $\Omega_{\mu\text{w}}/\sqrt{2}$ will cause transitions between the dressed-states. This feature has been used to create an effective clock qubit out of the combination of $\ket{D}$ with the state $\ket{0'}$, which does not form part of the dressed-state system.
Such a qubit has been shown to exhibit a significant increase in coherence time compared to magnetic field sensitive bare-state qubits and is a promising approach to microwave based quantum computing \cite{Timoney, Webster}.
The states \ket{u} and \ket{d} are also protected against decoherence by magnetic field fluctuations in the same way as \ket{D} and the set of states \ket{D}, \ket{u} and \ket{d} can be used to embody a qutrit. Unlike \ket{D} however, \ket{u} and \ket{d} are susceptible to decoherence caused by fluctuations in the power of the microwave dressing fields, potentially reducing the coherence time of the qutrit compared to the qubit.

The above analysis ignores the second-order effects of magnetic field fluctuations on the energies of the bare states. These second-order fluctuations are not decoupled by the dressing fields, but are small for $\chi\ll1$ and will limit possible coherence times to be similar to those obtained by the clock qubit.

\section{Preparation and detection of the dressed system}\label{sec:prepdet}

In order to realise the full potential of the dressed system, a method to prepare and detect all three dressed-states has been developed. Before preparing the dressed-states, the ion must first be initialised in a bare atomic state.
The ion is initially Doppler cooled on the 369nm near cycling $^{2}$S$_{1/2}\ket{F=1}$ $\leftrightarrow$ $^{2}$P$_{1/2}\ket{F=0}$ transition, with microwaves resonant with \ket{0}$\leftrightarrow$\ket{0'} repumping from F=0. To prepare the ion in $\ket{0}$ the microwaves are turned off and an EOM is used to modulate the 369nm light at 2.1GHz to address the $^{2}$S$_{1/2}\ket{F=1}$ $\leftrightarrow$ $^{2}$P$_{1/2}\ket{F=1}$ transition, with this light applied for 10 $\mu$s. Based on the 369nm laser intensity and modulation depth, we estimate a preparation infidelity of $<10^{-4}$. The final state is detected using a state-dependent fluorescence measurement by again applying 369 nm light resonant with the $^{2}$S$_{1/2}\ket{F=1}$ $\leftrightarrow$ $^{2}$P$_{1/2}\ket{F=0}$ cycling transition for 1.5 ms \cite{Webster}.
Scattered photons are then detected on a photomultiplier tube and a threshold is set to discriminate between a bright (fluorescing) and dark (not fluorescing) ion. 
Ideally the measurement would give a dark result if the ion is in $\ket{0}$ and a bright result if the ion is in any of the states $\ket{-1}$, $\ket{0'}$ or $\ket{+1}$ however in practice the procedure is imperfect due to overlap between the probability distributions of the number of photons detected for each case. We measure the conditional probabilities p(bright$\mid$F=0) and p(bright$\mid$F=1) then use this data to infer p(F=0) or p(F=1) from measurements of p(bright).
Since both the preparation and detection processes operate in the bare-state basis, an appropriate transfer sequence is required in order to prepare and detect the three dressed-states.

Previous works \cite{Timoney,Webster} have demonstrated preparing and detecting the dressed-state $\ket{D}$ by adiabatically ramping the amplitudes of the microwave dressing fields in a STIRAP process. Our new method, involving short pulses on the clock transition, reduces the experimental complexity and allows for all three dressed-states to be prepared and detected. We will begin by giving a brief summary of the previously used STIRAP method. We will then outline our new method and demonstrate the efficient preparation and readout of all three dressed-states.

\begin{figure}
\includegraphics [width=70mm]{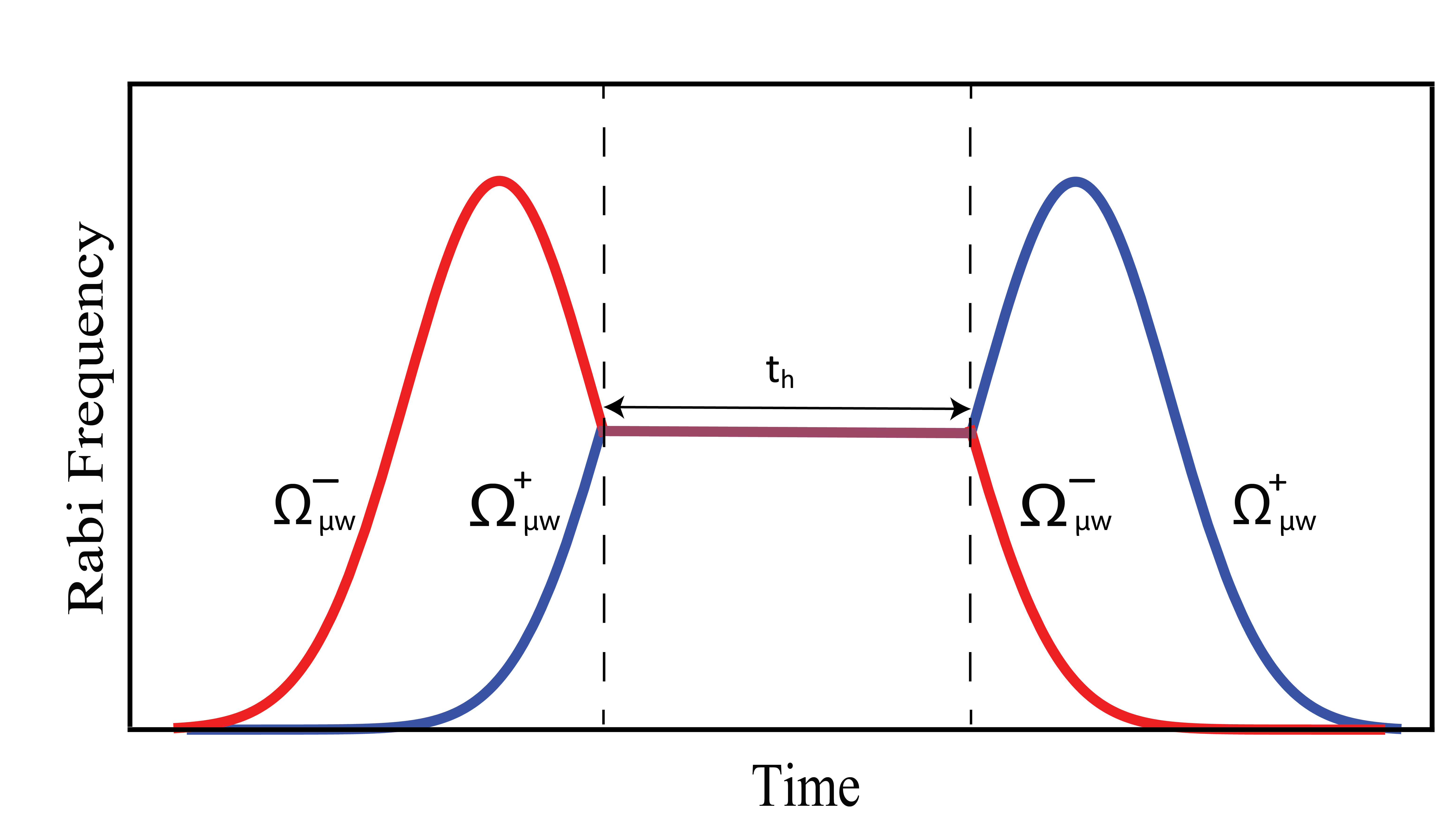}
\caption{Illustration of the STIRAP process. The microwave fields are ramped adiabatically in a particular order that transfers population from $\ket{+1}$ to the dressed-state $\ket{D}$. The fields are then held at equal Rabi frequencies for a hold time $t_h$, during which coherent manipulation in the dressed basis can be performed. Finally, the fields are ramped down, transferring any population in $\ket{D}$ to $\ket{-1}$.}
\label{STIRAP}
\end{figure}

\begin{figure*}
\includegraphics [width=160mm]{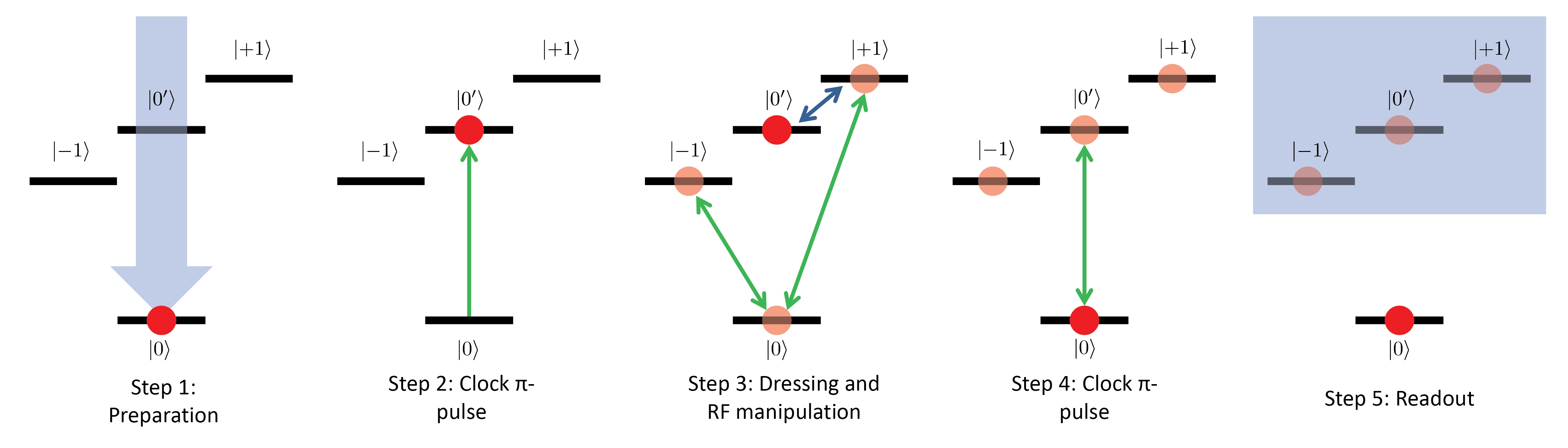}
\caption{Diagram illustrating the transfer sequence to prepare and detect the dressed-states via the clock transition. After the ion is prepared in $\ket{0}$ by optical pumping, a $\pi$-pulse on the clock transition transfers population to $\ket{0'}$ The dressing fields are then turned on and coherent manipulation in the dressed basis is performed using radio frequency fields (light red circles indicate population in the dressed-states). The dressing fields are then turned off and a second $\pi$-pulse on the clock transition transfers population from $\ket{0'}$ to $\ket{0}$. Any population that had been transferred to the dressed-states in step 3 will therefore give a bright result when reading out, while population in $\ket{0'}$ will give a dark result.}
\label{clock_method_diagram}
\end{figure*}

For the STIRAP method for preparing and detecting the dressed-state $\ket{D}$, starting from the initial $\ket{0}$ state we apply a $\pi$-pulse to transfer the population to $\ket{+1}$. The amplitudes of the microwave dressing fields are then slowly varied in time with Gaussian pulse shapes. For microwave dressing field Rabi frequencies $\Omega_{\mu\text{w}}^+$ and $\Omega_{\mu\text{w}}^-$, a STIRAP amplitude profile as shown in Fig. \ref{STIRAP} adiabatically transfers population between $\ket{+1}$ and $\ket{-1}$. When the amplitudes of the two fields are equal, population initialised in state $\ket{+1}$ has been transferred to the dressed-state $\ket{D}$. The dressing fields can then be held constant for a time $t_{h}$ for further experiments to be performed. The amplitudes are then ramped down to map any population in $\ket{D}$ to the bare-state $\ket{-1}$. To complete the transfer procedure, a microwave $\pi$-pulse after the STIRAP pulse sequence transfers population from $\ket{-1}$ to $\ket{0}$. A measurement therefore gives a dark result if the ion was in the dressed-state $\ket{D}$, and a bright result if the ion was in the dressed-states $\ket{u}$ or $\ket{d}$ or the bare-state $\ket{0'}$.  

This method does not easily lend itself to preparing and detecting all three dressed-states, which would be essential for the realisation of experiments involving qutrits \cite{Cohen, Klimov, Ralph, Collins, Kaszlikowski, Bartlett, Lanyon2, Senko, Bruss, Cerf} in a system that is well protected from magnetic noise. Furthermore, microwave based experiments often require a static magnetic field gradient, which causes the transition frequencies to vary between ions (see Eq. \ref{breitrabi}). The result is that each ion requires a pair of microwave dressing fields. Amplitude modulating these fields is possible yet undesirable as it complicates the experimental setup, especially when scaling to a large number of ions. Additionally, the requirement to prepare and detect a dressed-state via a STIRAP process involving a magnetic field sensitive transition can cause decoherence. It is therefore beneficial to develop a new method which mitigates these limitations.

Our new method for preparing and detecting the dressed-states utilises the clock transition $\ket{0} \leftrightarrow \ket{0'}$. By applying a microwave field resonant with this transition, a $\pi$-pulse will transfer population initialised in $\ket{0}$ to $\ket{0'}$, as shown in Fig. \ref{clock_method_diagram}. Since $\ket{0'}$ is not part of the dressed system, the microwave dressing fields can be turned on instantaneously without affecting the populations. At this point, if we are intending to use the states \ket{0'} and \ket{D} as a qubit, then the initialisation process is complete.

To transfer the ion from \ket{0'} to one of the dressed states we now apply a single RF field tuned near to either the $\ket{0'} \leftrightarrow \ket{+1}$ or $\ket{0'} \leftrightarrow \ket{-1}$ transition. In the interaction picture with respect to the atomic Hamiltonian and after making the RWA, the Hamiltonian of the RF field in the bare-state basis is given by

\begin{equation}\label{barerfH}\begin{split}
\hat{H}_\text{rf} = \frac{\hbar\Omega_\text{rf}}{2}\bigg(\ket{+1}\bra{0'}e^{-i\Delta_+t}
+ \ket{-1}\bra{0'}e^{i\Delta_-t} + \text{H.c}\bigg),
\end{split}\end{equation}
where we have set the phase to zero for clarity and $\Delta_+ = \omega_\text{rf} - \omega_B^+$ ($\Delta_- = \omega_\text{rf} - \omega_B^-$) is the detuning of the field from the $\ket{0'} \rightarrow \ket{+1}$ ($\ket{0'} \rightarrow \ket{-1}$) transition. The difference between $\Delta_+$ and $\Delta_-$ is fixed by the magnetic field to be $\Delta_+ -\Delta_- = \omega_B^- - \omega_B^+$ and can be calculated using Eq. \ref{breitrabi}. Eq. \ref{barerfH} can be written in the dressed basis as

\begin{equation}\begin{split}
\hat{H}_{\text{rf}}& = \frac{\hbar\Omega_{\text{rf}}}{2\sqrt{2}}\bigg(\ket{D}\bra{0'}\big(e^{-i\Delta_+t}- e^{i\Delta_-t}\big)+ \text{H.c}\bigg)\\
&+ \frac{\hbar\Omega_{\text{rf}}}{4}\bigg(\big(\ket{u}+\ket{d}\big)\bra{0'}\big(e^{-i\Delta_+t}+ e^{i\Delta_-t}\big) + \text{H.c}\bigg).
\end{split}\end{equation}
We can now transform to the interaction picture with respect to the dressing Hamiltonian (Eq. \ref{dressed_mw_hamiltonian}) to get

\begin{equation}\label{rfH}\begin{split}
\hat{H}'_{\text{rf}} = &\frac{\hbar\Omega_{\text{rf}}}{2\sqrt{2}}\bigg(\ket{D}\bra{0'}\big(e^{-i\Delta_+t}- e^{i\Delta_-t}\big) + \text{H.c}\bigg) \\
&+ \frac{\hbar\Omega_{\text{rf}}}{4}\bigg(\ket{u}\bra{0'}\big(e^{-i(\Delta_+-\frac{\Omega_{\text{$\mu$w}}}{\sqrt{2}})t} + e^{i(\Delta_-+\frac{\Omega_{\text{$\mu$w}}}{\sqrt{2}})t} \big) \\
&+ \ket{d}\bra{0'}\big(e^{-i(\Delta_++\frac{\Omega_{\text{$\mu$w}}}{\sqrt{2}})t} + e^{i(\Delta_--\frac{\Omega_{\text{$\mu$w}}}{\sqrt{2}})t} \big) +\text{H.c}\bigg).
\end{split}\end{equation}
This Hamiltonian contains six possible transitions from $\ket{0'}$, which can be selected by choosing the appropriate frequency for the RF field. Setting $\Delta_+ = 0$ or $\Delta_- = 0$ results in a transition between $\ket{0'}$ and the dressed-state $\ket{D}$ with Rabi frequency $\Omega_{\text{rf}}/\sqrt{2}$. Similarly, a detuning of $\Delta_+ = \Omega_{\text{$\mu$w}}/\sqrt{2}$ or $\Delta_- = -\Omega_{\text{$\mu$w}}/\sqrt{2}$ gives a transition between $\ket{0'}$ and $\ket{u}$ and for $\Delta_+ = -\Omega_{\text{$\mu$w}}/\sqrt{2}$ or $\Delta_- = \Omega_{\text{$\mu$w}}/\sqrt{2}$, population is transferred between $\ket{0'}$ and $\ket{d}$, both with Rabi frequency $\Omega_{\text{rf}}/2$. Each of the three dressed-states can therefore be prepared by applying an RF $\pi$-pulse resonant with one of these transitions.

The phases of the RF fields can be changed as would be normal for a driven two level system to allow different rotations of the states. Note that since there are two RF transitions from \ket{0'} to each of the dressed states, the relative phases of the RF and microwave fields do become important if both are to be used to manipulate the state.

To detect if population has been transferred to the desired dressed-state, the dressing fields can be turned off instantaneously. Population in $\ket{0'}$ will be unaffected, and a second microwave $\pi$-pulse resonant with the clock transition will swap population between $\ket{0'}$ and $\ket{0}$. Any population that had been transferred to the dressed-states by the RF field will now be in the $\{\ket{-1},\ket{0'},\ket{+1}\}$ manifold and will therefore give a bright result upon detection, whereas population that remained in $\ket{0'}$ will now be in $\ket{0}$ and will therefore give a dark result.

\begin{figure}
\includegraphics [width=75mm]{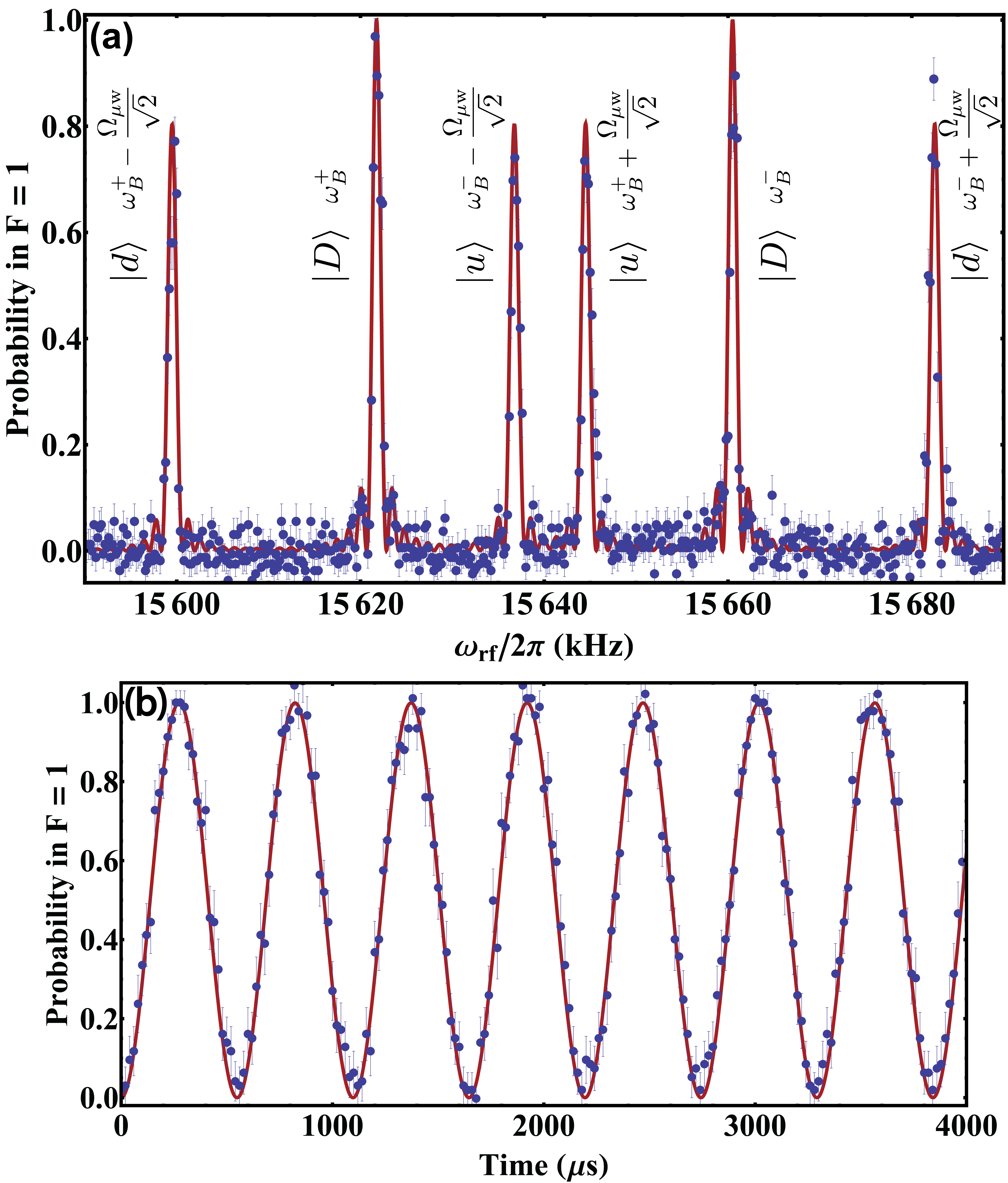}
\caption{(a) Population in F = 1 after a frequency scan of a single RF field applied for 800 $\mu$s, after preparing the ion in $\ket{0'}$ and applying the microwave dressing fields. The red line is a theory curve describing the sum of six transition probabilities, with the six center frequencies and the Rabi frequency of the $\ket{0'} \leftrightarrow \ket{D}$ transition, $\Omega_D$, as free parameters. The Rabi frequency is determined to be $\Omega_D/2\pi = 0.63$ kHz and the difference in peak height (corresponding to the difference in Rabi frequencies between the transitions) between $\ket{D}$ and $\ket{u}$ and $\ket{d}$ has been fixed according to equation 9. From the transition frequencies, we extract the microwave Rabi frequency, which determines the separation between $\ket{D}$, $\ket{u}$ and $\ket{d}$ (labelled in figure), to be $\Omega_{\text{$\mu$w}}/\sqrt{2} = 2\pi \times 23(1)$ kHz, and the second-order Zeeman shift, which determines the separation between the two sets of transitions, to be $|\omega_B^- - \omega_B^+|/2\pi = 39(1)$ kHz. The main source of error in this measurement is attributed to uncompensated drifts in the B-field which occur during the long period of data acquisition. (b) Rabi oscillations between $\ket{0'}$ and $\ket{D}$. Population is prepared in $\ket{0'}$ via a resonant microwave $\pi$-pulse on the $\ket{0} \leftrightarrow \ket{0'}$ transition with a duration of 14 $\mu$s. The red line represents a theory curve describing a resonant Rabi oscillation. The free parameters are the Rabi frequency and the contrast, which were determined to be $\Omega_D/2\pi = 1.8$ kHz and 0.99(1).}
\label{dressedscan}
\end{figure}

To demonstrate our new method we use a single $^{171}$Yb$^+$ ion trapped in a linear Paul trap \cite{McLoughlin2} and apply a static magnetic field of approximately 11 Gauss, for which we have measured $\omega_B^+/2\pi =15.622$ MHz, $\omega_B^-/2\pi =15.661$ MHz and $|\omega_B^- -\omega_B^+|/2\pi= 39$ kHz. Following preparation in $\ket{0}$ we transfer population to $\ket{0'}$ using a resonant microwave $\pi$-pulse. We then instantaneously apply the microwave dressing fields, where $\Omega_{\text{$\mu$w}}/2\pi = 31$ kHz, followed by manipulation with a single RF field to prepare one of the dressed-states. To read out the final state after manipulation, the dressing fields are turned off and a second $\pi$-pulse is applied on the clock transition. Fig. \ref{dressedscan} (a) shows the population in F = 1 after a frequency scan of a single RF field that has been applied for 800 $\mu$s, corresponding to the measured $\pi$-time of the $\ket{0'} \leftrightarrow \ket{D}$ transition, using our new method for preparing and detecting. The six peaks correspond to the six transitions in Eq. \ref{rfH}, which indicate transitions to the three dressed-states $\ket{d}$, $\ket{D}$ and $\ket{u}$ via the $\ket{0'}\leftrightarrow\ket{+1}$ or the $\ket{0'}\leftrightarrow\ket{-1}$ transition. The peaks vary in height due to the different Rabi frequencies for the transitions. As an example we set the RF frequency to $\Delta_+ = 0$ to induce Rabi oscillations between $\ket{0'}$ and $\ket{D}$ with Rabi frequency $\Omega_{\text{rf}}/\sqrt{2} = 2\pi \times 1.8$ kHz, as shown in Fig. \ref{dressedscan} b. In this case the Rabi frequency of the microwave dressing fields has been set to $\Omega_{\text{$\mu$w}}/2\pi = 29$ kHz and the preparation and detection $\pi$-pulses were performed in 14 $\mu$s.

\begin{figure}
\includegraphics [width=75mm]{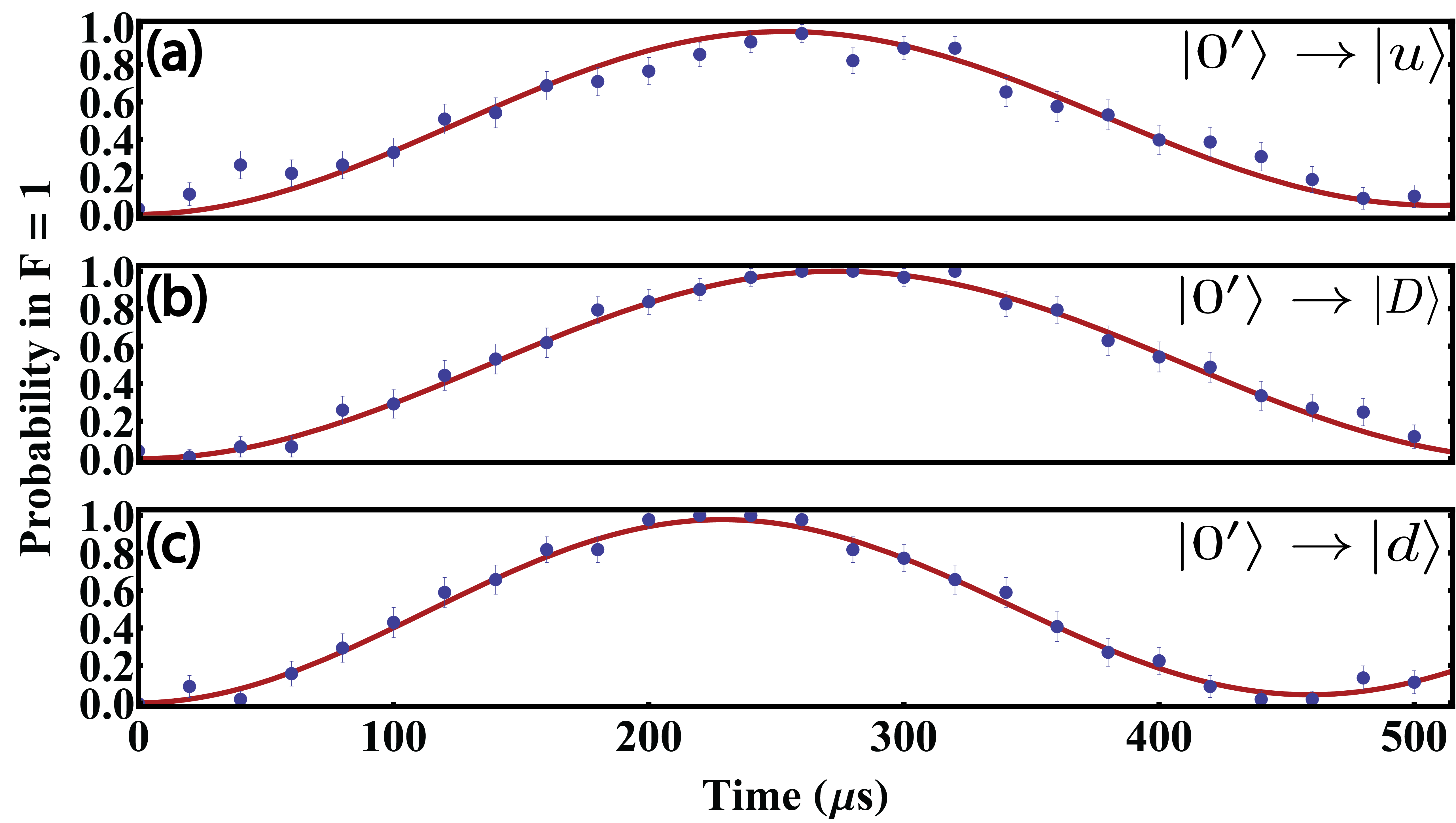}
\caption{(a), (b) and (c) show a Rabi oscillation between $\ket{0'}$ and $\ket{u}$, $\ket{D}$ and $\ket{d}$ respectively. The microwave dressing field Rabi frequencies were set to $\Omega_{\text{$\mu$w}}/2\pi = 29$ kHz for (a) and (b) and $\Omega_{\text{$\mu$w}}/2\pi = 47$ kHz for (c). The dressed-state Rabi frequencies were measured to be $\{\Omega_u,\Omega_D,\Omega_d\}/2\pi = \{2.0, 1.8, 2.2\}$ kHz. Note that different RF powers were used in each case.}
\label{dressed_flops}
\end{figure}

Figs. \ref{dressed_flops}(a), \ref{dressed_flops}(b) and \ref{dressed_flops}(c) show a Rabi oscillation between $\ket{0'}$ and each of the three dressed-states $\ket{u}$, $\ket{D}$ and $\ket{d}$ respectively. We measure a preparation fidelity of 0.99(1) for $\ket{D}$ and 0.98(1) for $\ket{u}$ and $\ket{d}$. Increasing and stabilising the dressing field power is expected to further improve the preparation fidelities. We have also measured the lifetimes to be 700 ms for $\ket{D}$ and 70 ms for $\ket{d}$, both of which are far greater than typical interaction times. Our setup is not optimised to minimise microwave amplitude noise, therefore we anticipate significant improvements could be made for the lifetimes of \ket{d} and \ket{u} which are limited by uncompensated amplitude fluctuations of the microwave driving field.

There are several conditions that have to be fulfilled to ensure that multiple transitions are not driven simultaneously. Firstly, the Rabi frequency of the RF field should satisfy $\Omega_{\text{rf}} \ll \big(|\omega_B^- - \omega_B^+|$, $\Omega_{\text{$\mu$w}}\big)$ so that both Zeeman levels are not driven at the same time and the energy gap to $\ket{u}$ and $\ket{d}$ is not bridged. Secondly, it should be ensured that none of the transitions overlap, for example if $|\omega_B^- - \omega_B^+| \approx \Omega_{\text{$\mu$w}}/\sqrt{2}$, transitions to $\ket{u}$ and $\ket{D}$ would have the same resonant frequency in both instances. This would prohibit the individual preparation of $\ket{u}$ and $\ket{D}$. In practice, these conditions can be satisfied by ensuring that all six peaks can be clearly resolved in an experiment such as shown in Fig. \ref{dressedscan} (a). Numerical simulations for the parameters in this case indicate that, absent any decoherence, the infidelity due to these other transitions is $6\times10^{-4}$ for a $\pi$ pulse from \ket{0'} to \ket{D}. This can be reduced by changing parameters - for instance reducing $\Omega_{\rm rf}$ by a factor of 10 reduces this infidelity to $7\times10^{-6}$.

With further RF manipulation, the method presented here can lead to the full manipulation and characterisation of a qutrit encoded in the three-level dressed system. Manipulation of the qutrit can be performed using multiple RF pulses on each of the transitions between $\ket{0'}$ and $\ket{d}$, $\ket{D}$, $\ket{u}$. As an example, a $\pi$-pulse from $\ket{d}$ to $\ket{0'}$, followed by a $\pi/2$-pulse on the $\ket{0'} \leftrightarrow \ket{u}$ transition and finally a second $\pi$-pulse from $\ket{0'}$ to $\ket{d}$ would effectively perform a $\pi/2$ rotation between $\ket{u}$ and $\ket{d}$. To read out population in one of the three dressed-states, an RF $\pi$-pulse can be used to map the population to $\ket{0'}$, after which a clock $\pi$-pulse can be used to transfer population to $\ket{0}$, allowing for the state to be detected. Therefore with RF and microwave manipulation and subsequent measurements, an arbitrary qutrit state could be reconstructed using quantum state tomography \cite{James2}.

\section{Individual addressing of a clock transition}\label{sec:clock}

It would be advantageous to combine the preparation and detection method demonstrated in the previous section with a static magnetic field gradient \cite{Mintert1} to perform multi-ion quantum operations using microwave dressed-states. To achieve this, the effects of adding a strong magnetic field gradient should be considered. For multiple ions in a magnetic field gradient, transitions between magnetic field sensitive states in different ions are separated in frequency space, allowing for individual addressing of ions in a string \cite{Mintert1,Johanning2}. Individual ion addressing with fault tolerant cross-talk values on the order of $10^{-5}$ has been achieved by C. Piltz et al. \cite{Piltz2}. For a large gradient, there is also a significant difference between the clock transition frequencies in different ions due to the second order Zeeman shift, and this difference has to be taken into account when using the clock method to prepare and detect the dressed-states in each ion. By ensuring that the Rabi frequency of the microwave field addressing the clock transition in each ion is much less than the splitting between them, the transitions can be addressed individually. The use of one microwave field per ion while obeying this condition therefore allows for the preparation and detection method to be combined with the static magnetic field gradient scheme. Furthermore, the ability to individually address these transitions would allow ions in a string to be selected for preparation in the dressed basis, while other ions would be unaffected.

For $N$ ions in a trap with axial secular frequency $\nu$, the splitting between clock transitions in ions $i$ and $j$ can be calculated using Eq. \ref{breitrabi} to be

\begin{equation}\label{eq:nonlinear}
\Delta\omega_{B,ij}^0 = |\omega_B^0(B_i) - \omega_B^0(B_i + d_{ij}\partial_zB)|,
\end{equation} 
where $B_i$ is the magnetic field at ion $i$, $d_{ij} \propto 1/\nu^{2/3}$ is the distance between ions $i$ and $j$ and $\partial_zB$ is the axial magnetic field gradient, which is assumed to be equal for both ions.

\begin{figure}
\includegraphics[width=75mm]{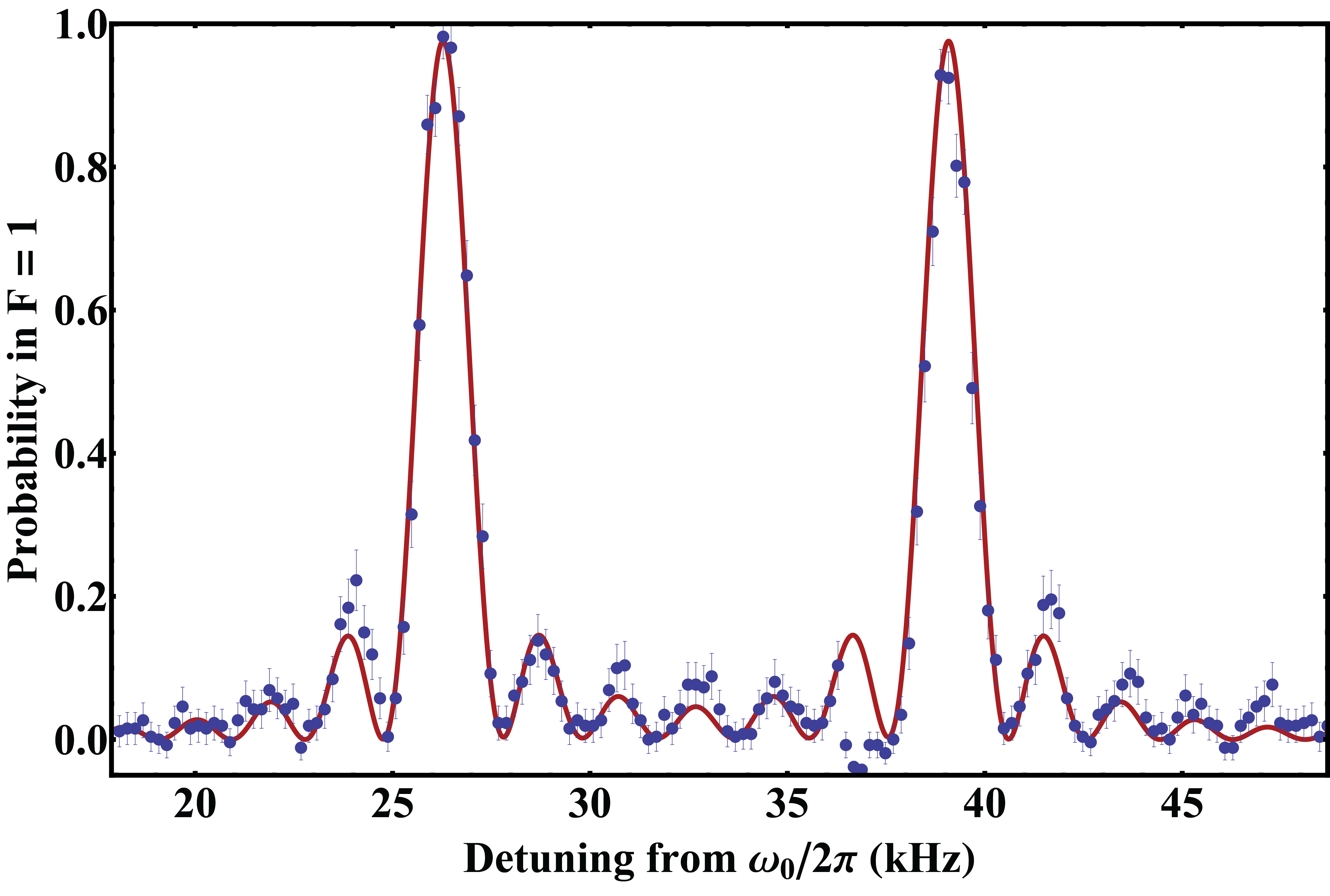}
\caption{Population in F = 1 after a frequency scan over the $\ket{0} \leftrightarrow \ket{0'}$ clock transition for two ions using a single microwave field. The red line is a theory curve describing the sum of two transition probabilities, with the Rabi frequency fixed at $\Omega_{\text{clock}}/2\pi = 0.9$ kHz and the two transition frequencies as free parameters. From this we determine the separation between the two peaks due to the second order Zeeman effect to be $\Delta\omega_B^0/2\pi = 12.8$ kHz.}
\label{individualaddressingclock}
\end{figure}

We have designed and built an experiment that combines a linear Paul trap with four Samarium Cobalt permanent magnets, resulting in an axial magnetic field gradient of $\partial_zB = 23.6(1)$ T/m at the position of the ion string \cite{Lake}. With this gradient, an offset magnetic field of $B_1 \approx 9$ Gauss and an axial secular frequency of $\nu/2\pi = 268$ kHz, we find a frequency separation between the clock transitions in two ions of $\Delta\omega_{B}^0/2\pi = 12.8 $ kHz. Fig. \ref{individualaddressingclock} shows the population in F = 1 after a frequency scan over the $\ket{0} \leftrightarrow \ket{0'}$ transition for two ions using a single microwave field. To ensure that there is minimal cross-talk between the transitions, the Rabi frequency for this transition, $\Omega_{\text{clock}}$, satisfied the condition $\Omega_{\text{clock}} \ll |\omega_B^- - \omega_B^+|$. In this case the $\pi$-pulse time was set to 550 $\mu$s, resulting in a cross-talk value of $\Omega_{\text{clock}}^2/|\omega_B^- - \omega_B^+|^2 \approx 5.0 \times 10^{-3}$, which indicates the fractional excitation of one transition when resonantly driving another that is separated in frequency \cite{Piltz2}. The reduction in speed compared to the single ion case is not a significant problem as the $\pi$-pulse time is still much shorter than the coherence time for this transition, which can exceed 1 s \cite{Fisk}. To further reduce the $\pi$-pulse time and the cross-talk, the magnetic field gradient or magnetic field offset ($B_1$) could be increased \footnote{Note that increasing $B_1$ will increase the sensitivity of the transition to magnetic field noise; cross-talk scales with $1/B_1^2$ and sensitivity with $B_1$}, or the secular frequency could be lowered as can be inferred from Eq. \ref{eq:nonlinear}. 

Individual addressability of clock transitions could also be useful in laser-based schemes as an alternative to tightly focussed laser beams for individual qubit addressing \cite{Nagerl2}. For example, with two global counter-propagating Raman beams equally illuminating an ion string in a strong magnetic field gradient, entanglement gates between non-nearest-neighbouring ions could be performed using clock qubits \cite{Haljan2}. Making use of the individually tunable interaction strength between pairs of ions, this could also provide a new method for the realisation of quantum simulations of spin models on an arbitrary lattice \cite{Korenblit}.
 \\

\section{Conclusion}

We have developed and implemented a new method for preparing and detecting all three states of a three-level dressed system. This method greatly simplifies the experimental setup compared to previous methods, which could lead to higher dressed-state preparation and detection fidelities and help to implement high-fidelity multi-ion entanglement gates with dressed-state qubits. Furthermore, our method allows all three of the dressed-states to be prepared and detected, providing access to a qutrit which is well protected from magnetic field noise. This system should allow the implementation of experiments involving qutrits such as discussed in Refs. \cite{Cohen, Klimov, Ralph, Collins, Kaszlikowski, Bartlett, Lanyon2, Senko, Bruss, Cerf}. We have shown that our method can be combined with a static magnetic field gradient by individually addressing the clock transitions in two ions, and therefore demonstrating that multiple ions in a string can be prepared and detected in the dressed basis with little cross-talk. This opens up the possibility of implementing high fidelity multi-ion quantum computation and simulation with two and three-level systems using microwaves. Furthermore, the ability to individually address clock qubits would allow the individual control of pairwise interaction strengths between arbitrary ions in a string using lasers. 

\section{Acknowledgements}

This work is supported by the U.K. Engineering and
Physical Sciences Research Council (EP/E011136/1,
EP/G007276/1), the European Commission’s Seventh
Framework Programme (FP7/2007-2013) under Grant
Agreement No. 270843 (iQIT), the Army Research
Laboratory under Cooperative Agreement No. W911NF-
12-2-0072 and W911NF-14-2-0106 and the University of Sussex. The views and
conclusions contained in this document are those of the
authors and should not be interpreted as representing the official policies, either expressed or implied, of the Army Research Laboratory or the U.S. Government. The U.S. Government is authorized to reproduce and distribute reprints for Government purposes notwithstanding any copyright notation herein.

\bibliography{ReportBib}

\end{document}